\documentclass[letterpaper]{mc2023}
\usepackage[caption=false]{subfig}
\usepackage{float}
\usepackage{tabls}
\usepackage{cites}
\usepackage{epsf}
\usepackage{appendix}
\usepackage{ragged2e}
\usepackage[top=1in, bottom=1in, left=1in, right=1in]{geometry}
\usepackage{enumitem}
\usepackage{xcolor}

\setlist[itemize]{leftmargin=*}
\usepackage{caption}
\title{Development of MC/DC: a performant, scalable, and portable\\Python-based Monte Carlo neutron transport code}
\author{%
  \textbf{Ilham Variansyah$^1$, J. P. Morgan$^2$, Jordan Northrop$^1$,}\\
  \textbf{Kyle E. Niemeyer$^2$, and Ryan G. McClarren$^3$}\vspace{3pt} \\
  \vspace{6pt}\\
  $^1$School of Nuclear Science and Engineering\\
  Oregon State University, Merryfield Hall, Corvallis, OR 97331 
  \vspace{6pt}\\
  $^2$School of Mechanical, Industrial, and Manufacturing Engineering\\
  Oregon State University, Rogers Hall, Corvallis, OR 97331 
  \vspace{6pt}\\
  $^3$Department of Aerospace and Mechanical Engineering\\
  University of Notre Dame,
  Fitzpatrick Hall, Notre Dame, IN 46556 \vspace{6pt}\\ 
  \url{variansi@oregonstate.edu}, \url{morgjack@oregonstate.edu}
}
\newcommand{\authorHead}{Variansyah, et al.}
\newcommand{\shortTitle}{Development of MC/DC}
\begin{document}
\maketitle
\justify 
\parskip 6pt plus 1 pt minus 1 pt

\begin{abstract}
We discuss the current development of MC/DC (Monte Carlo Dynamic Code). MC/DC is primarily designed to serve as an exploratory Python-based MC transport code. However, it seeks to offer improved performance, massive scalability, and backend portability by leveraging Python code-generation libraries and implementing an innovative abstraction strategy and compilation scheme.
Here, we verify MC/DC capabilities and perform an initial performance assessment. We found that MC/DC can run hundreds of times faster than its pure Python mode and about 2.5 times slower, but with comparable parallel scaling, than the high-performance MC code Shift for simple problems.
Finally, to further exercise MC/DC's time-dependent MC transport capabilities, we propose a challenge problem based on the C5G7-TD benchmark model.
\end{abstract}
\vspace{6pt}

\section{INTRODUCTION} 
The Center for Exascale Monte Carlo Neutron Transport (CEMeNT)\footnote{https://cement-psaap.github.io/} focuses on advancing the state of the art of Monte Carlo (MC) neutronics calculations, particularly for solving transient transport problems on exascale computer architectures.
One of CEMeNT's approaches is to develop an open-source Python-based MC code---MC/DC (Monte Carlo Dynamic Code) \cite{zenodoMCDC}---which is specifically designed for method and algorithm R\&D on both CPU and GPU architectures.
Written in Python, MC/DC offers rapid prototyping of MC methods and algorithms.
However, MC/DC leverages Python code-generation libraries to boost its performance and achieve massive parallel scalability and backend portability.
Initially inspired by the petascale-tested computational fluid dynamics code PyFR~\cite{Witherden2014PyFR}, this type of development approach is targeted to divorce the computer science from the numerical algorithms, making methods development and testing easier on subject area experts, as well as allowing portability between computer architectures.

In this paper, we discuss the development of MC/DC.
Section~\ref{sec:software_engineering} discusses the software engineering approach implemented.
Section~\ref{sec:result} discusses current capabilities, recent verification efforts, and initial performance assessment results.
Finally, we summarize and discuss ongoing and future work in Section~\ref{sec:future}.

\section{SOFTWARE ENGINEERING APPROACH}
\label{sec:software_engineering}

MC/DC serves two purposes: (1) to be a Python-based tool for MC transport method and algorithm explorations and (2) as a meta-programming demonstration toward performant, scalable, and portable Python-based MC transport.
To achieve these, we leverage Python code-generation libraries and implement an abstraction strategy.

Based on our previous explorations on the feasibility of Python-based hardware acceleration and abstraction techniques (tested on both x86 CPUs and Nvidia GPUs) \cite{morgan2022}, we chose Numba~\cite{lamNumba2015} as the basis for MC/DC development, as it provides a good balance of performance and ease of use.
Numba is a Python package that uses a just-in-time (JIT) compilation scheme to convert and compile functions written in native Python using the LLVM Compiler to ``approach C-like speeds'' \cite{lamNumba2015}. 
Furthermore, Numba can implement CPU threading via OpenMP and compile to Nvidia GPUs, which is an important piece of our MC/DC kernel abstraction strategy.

\begin{figure}[htbp]
  \centering
  \includegraphics[width=0.7\textwidth]{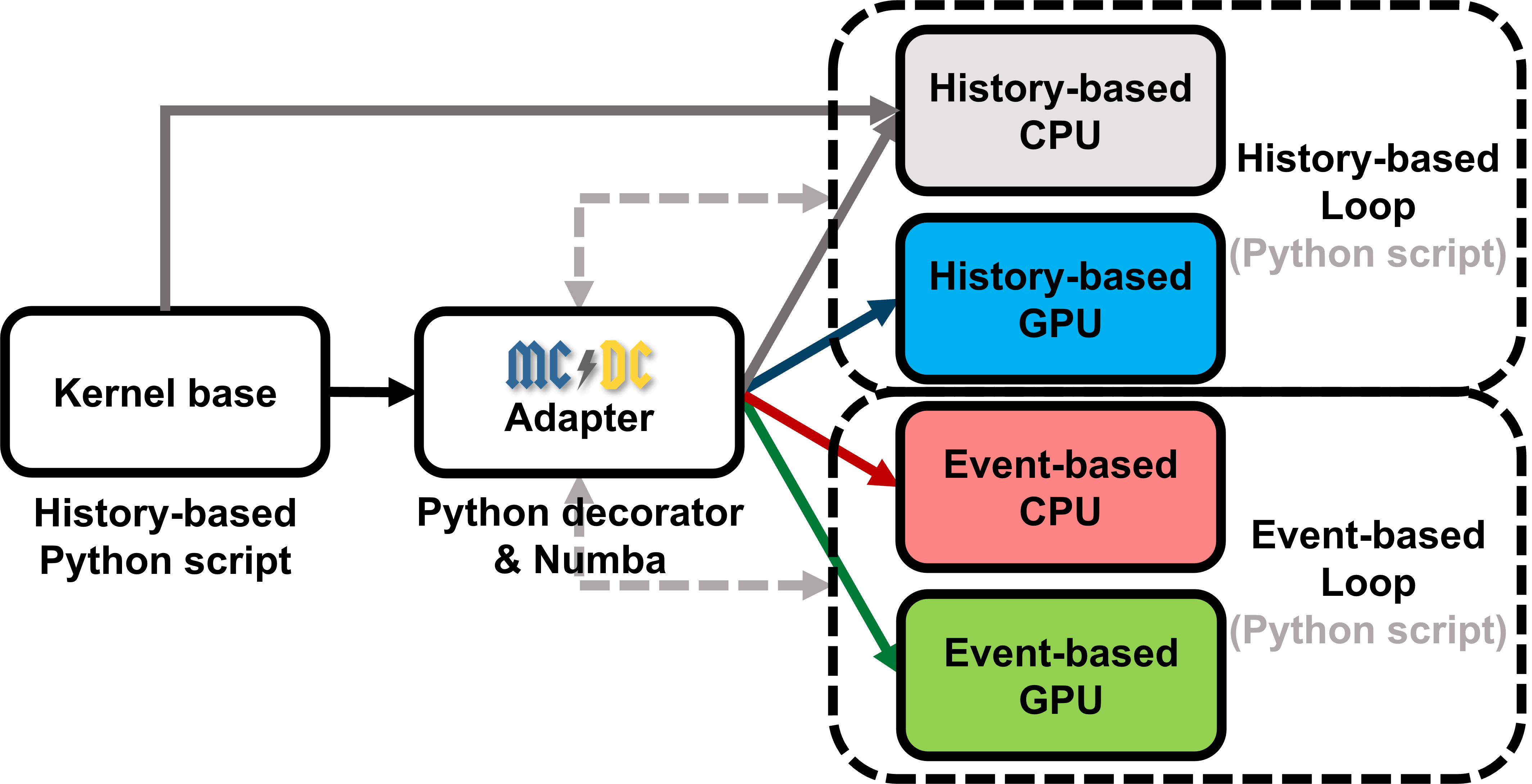}
  \caption{Current abstraction/portability strategy of MC/DC.}
  \label{fig:adapter}
\end{figure}

Figure~\ref{fig:adapter} illustrates the current abstraction strategy for MC/DC.
The objective is for all MC transport kernels (designed to act on a single particle on the typical history-based algorithm) to be strictly written in Python scripts.
Then adapters in MC/DC alter the behavior and compile kernels to run on one of the four targeted modes: combinations of history- or event-based algorithms with CPU or GPU backend targets.
These adapters are a series of Python decorators and Numba CPU/GPU JIT compilations.

\begin{figure}[htb]
  \centering
  \includegraphics[width=0.85\textwidth]{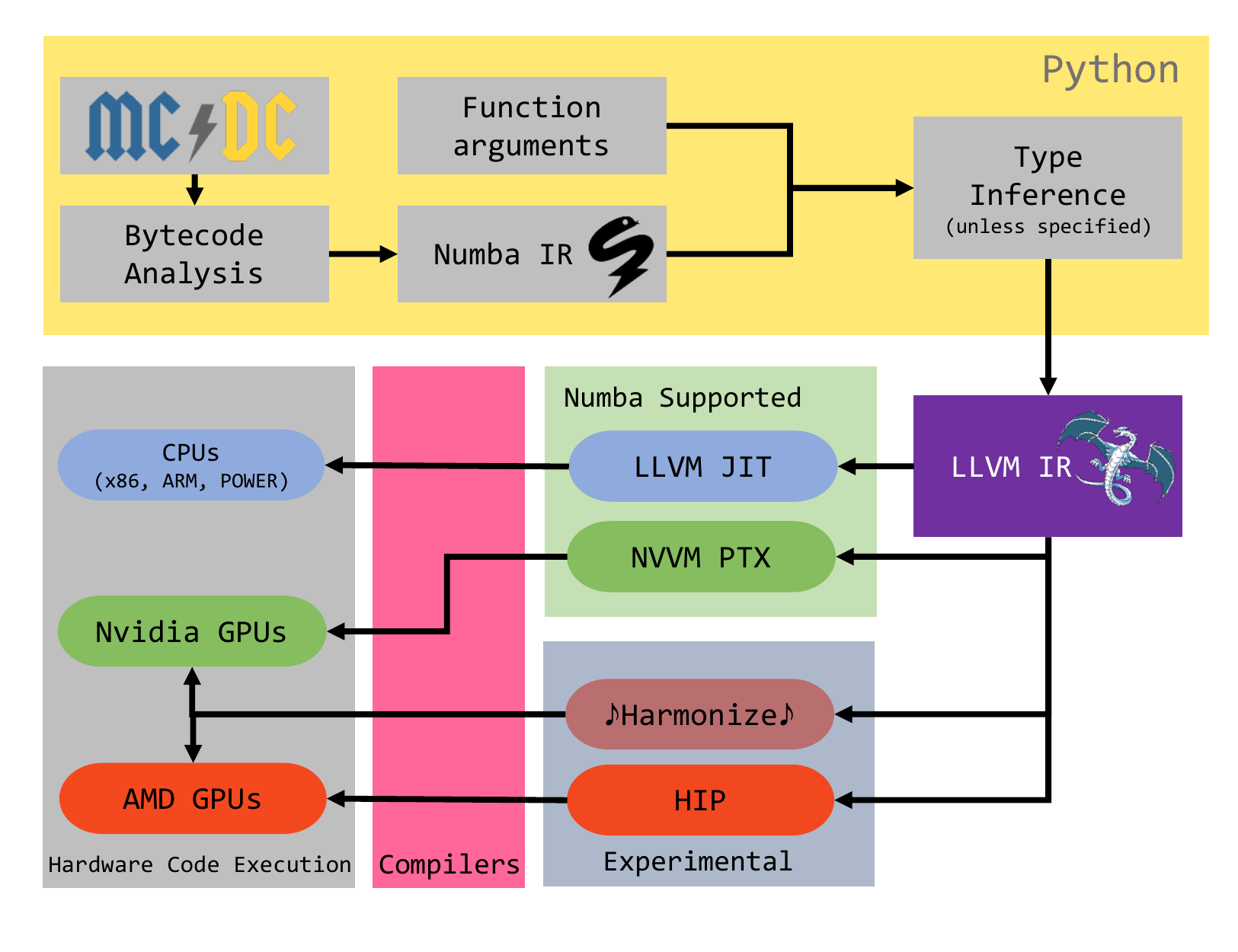}
  \caption{Proposed compilation path of abstracted MC/DC kernels.}
  \label{fig:compiler}
\end{figure}

Figure~\ref{fig:compiler} shows MC/DC's proposed compilation structure, including currently supported Numba functionality, planned extensions of our own fork of Numba to AMD GPUs, and integration of our own asynchronous GPU scheduler---Harmonize---for both AMD and Nvidia GPUs.
After MC/DC history-based MC transport kernels are wrapped and adapted, they are handed to Numba (shown in the yellow portion of Figure~\ref{fig:compiler}). 
From there, functions are lowered into the LLVM portability framework, then, in a currently supported path, compiled to Nvidia GPU and x86, ARM, and POWER CPU machine code.

Compiling kernels to AMD GPUs will be important for MC/DC since many exa-class machines use AMD GPUs as their primary accelerators (e.g., Oak Ridge's Frontier, LLNL's El Capitan). 
Numba does not currently support this.
However, since full LLVM does have support for AMD's Heterogeneous Interface for Portability (HIP) framework, we plan on elevating required functions in our own fork of Numba to gain compute access to AMD GPUs.
We hope that by using a compilation path reliant on LLVM---which seeks to be a ``source- and target-independent optimizer'' \cite{llvm}---our chosen acceleration and abstraction techniques will also allow us to readily extend MC/DC to other currently deployed accelerator hardware (e.g., Intel GPUs) and hardware yet to be designed.

Using LLVM intermediate representation (IR) will also allow us to extend MC/DC to use the Harmonize asynchronous GPU scheduler.
Harmonize works through a set of template abstractions and vendor-supplied compiler tools to queue GPU functional calls then run them in unison, thus decreasing thread divergence and increasing GPU performance.
It has been written specifically for transient Monte Carlo neutron transport with fission and provides 1.5$\times$ or better improvement in about 77.8\% of cases (varying material cross sections) in mono-energetic simulations \cite{cuneothesis}.
Further work is still required to demonstrate Harmonize's abilities in a more-complex tracking algorithm (e.g., multi-dimension, multigroup) and to extend its operability to work with LLVM IR (currently it works using Nvidia specific PTX code), but it shows promise for further increasing MC/DC's GPU performance without requiring significant alterations to algorithms.

Currently, MC/DC only runs with the history-based algorithm on CPU, with Numba-JIT compilation for performance improvement, and with MPI4Py~\cite{DALCIN20051108} for scalable parallel runs across compute nodes.
However, we have started investigating the proposed abstraction strategy (Fig.~\ref{fig:adapter}) and compilation path (Fig.~\ref{fig:compiler}) on a light version of MC/DC \cite{zenodoMCDC-BIT}.
In the next section, we discuss MC/DC's current capabilities, recent verification efforts, and initial performance assessment results for the history-based CPU mode.

\section{VERIFICATION AND PERFORMANCE ASSESSMENT}\label{sec:result}

MC/DC is capable of running fixed-source and eigenvalue neutron transport problems defined on quadric surface constructive solid geometry.
MC/DC has several time-dependent features, including time-dependent mesh tallies, time-dependent source, time census, population control techniques~\cite{variansyah2022pct,variansyah2022physor}, time-dependent surface for continuous object movement~\cite{variansyah2023mc_tdSurf}, and initial condition particle sampling for typical reactor transients~\cite{variansyah2023mc_ic}.

Currently, MC/DC only supports running multigroup transport problems; continuous-energy physics is planned to be implemented later in the project.
Starting with multigroup physics allows us to focus more on the main novel work of MC/DC development: Python-based transport kernel and algorithm abstractions, as well as investigating and exploring time-dependent MC techniques.
Furthermore, we do not see multigroup MC as a mere stepping stone to achieving the ultimate goal of running continuous-energy MC, as multigroup MC does have potential---it can be an excellent alternative, or complement, to the inherently multigroup, yet widely-used, deterministic codes.
With MC/DC, we want to explore optimal sets of MC algorithms and techniques for both multigroup and continuous-energy transports, which can be very different and highly problem dependent.

\begin{figure}[htbp]
    \centering
    \subfloat{%
    \resizebox*{7cm}{!}{\includegraphics{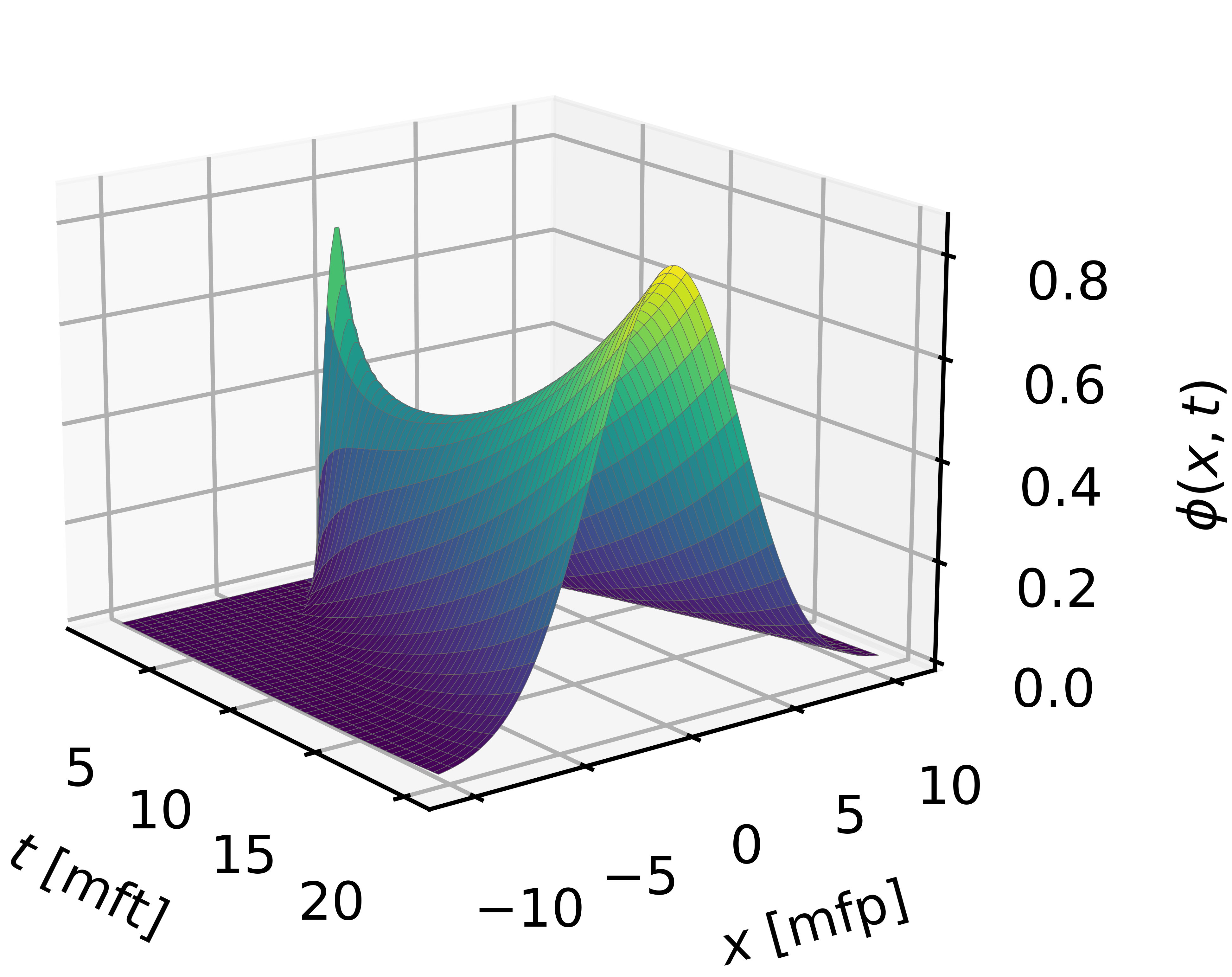}}}\hspace{8mm}
    \subfloat{%
    \resizebox*{7cm}{!}{\includegraphics{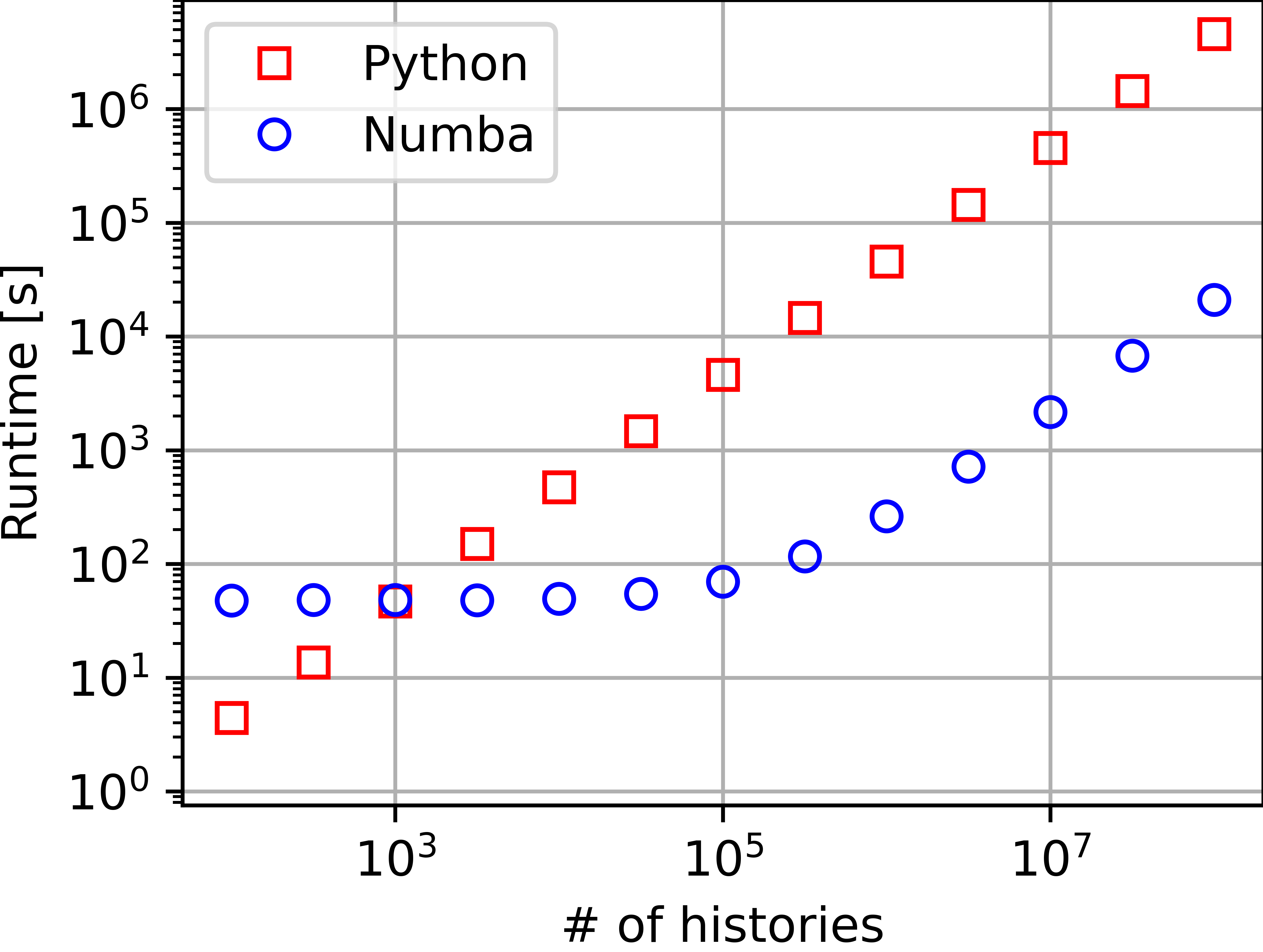}}}
    \caption{Solution (left) and Numba speedup (right) for the supercritical AZURV1.} \label{fig:azurv1}
\end{figure}

\begin{figure}[htbp]
  \centering
  \includegraphics[scale=0.7]{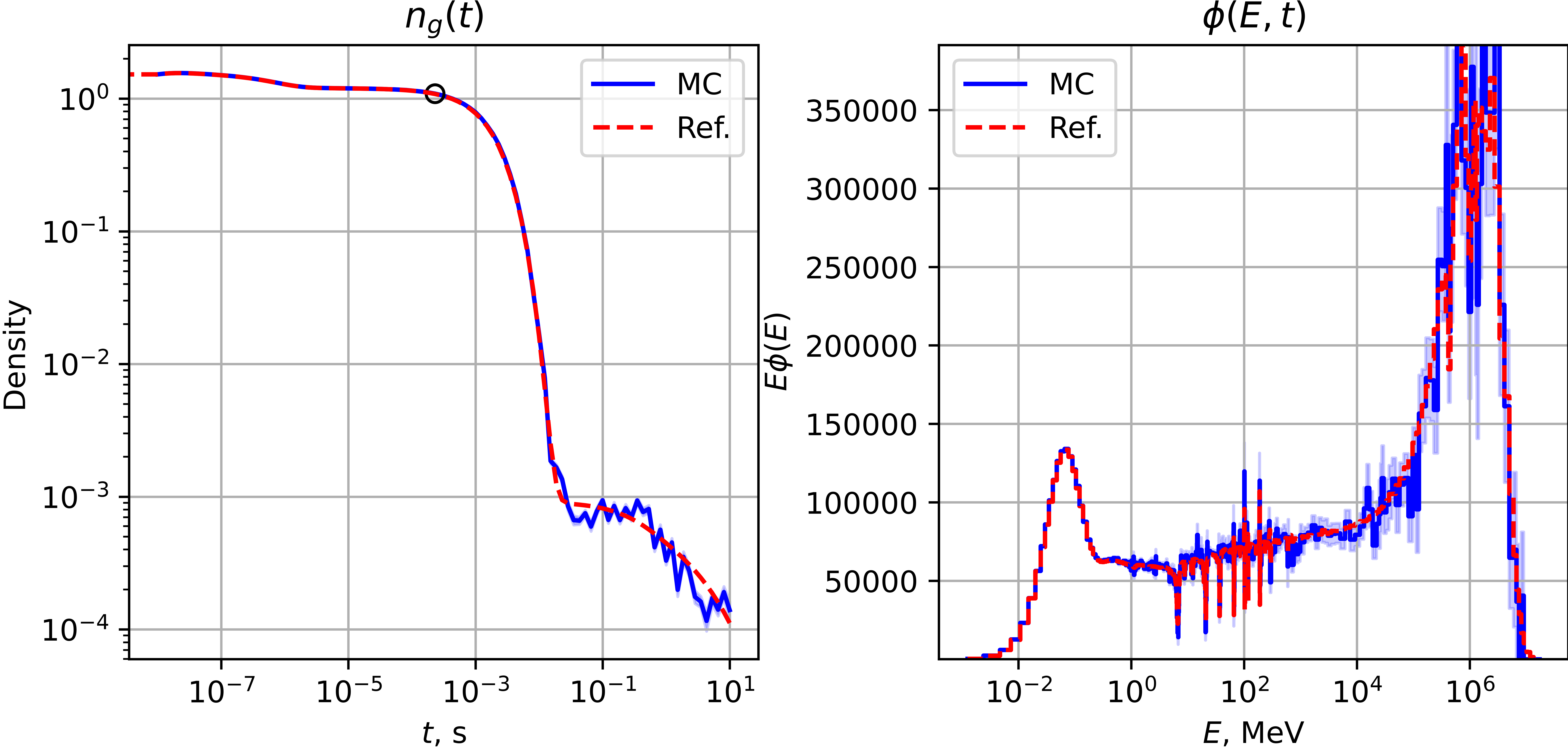}
  \caption{A solution snapshot of the solution of the neutron-pulse transient of an infinite subcritical homogeneous 361-group medium (10$^{\boldsymbol{6}}$ histories).}
  \label{fig:shem361}
\end{figure}

MC/DC capabilities have been verified against several analytical and numerical test problems.
These include the time-dependent supercritical AZURV1 \cite{Ganapol2001HomogeneousBenchmarks,variansyah2022pct}, where we add neutron fission to the original problem to achieve an effective multiplication ratio of $c=1.1$ to exercise a significant particle population growth, which is about 7.39 times of the initial value in 20 mean-free-time (mft). The semi-analytical solution for this problem is shown in the left figure of Fig.~\ref{fig:azurv1}. 
We also devised a neutron-pulse transient of a subcritical homogeneous 361-group medium, representing an infinite water reactor pin cell to verify the multigroup physics capability.
We tally the evolution of the multigroup flux in a logarithmically-spaced time grid with a time-crossing estimator~\cite{variansyah2022pct}.
Figure~\ref{fig:shem361} shows a snapshot of the result. 
Finally, part of MC/DC verification is observing the $1/\sqrt{N}$ convergence of the solution error~\cite{variansyah2022pct}, where $N$ is the number of histories, for problems where we know the analytical/accurate solutions, such as the supercritical AZURV1 and the 361-group pulse problems.

\subsection{Numba Speedup}

MC/DC can be run in pure Python or performant Numba mode. 
Pure Python mode allows rapid testing and development of new methods. 
Once an implementation is verified and ready for testing on more practical and computationally challenging problems, we can turn on the Numba mode, get the MC kernels JIT-compiled, and run performantly, approaching the speeds of compiled programs.
Figure~\ref{fig:azurv1} (right) demonstrates this, showing the Numba speedup by comparing the respective runtimes of Python and Numba modes for the supercritical AZURV1 problem.
The flat runtime for Numba mode on smaller numbers of histories indicates the JIT compilation time of about 45 seconds.
Running with sufficiently high numbers of histories hides the compilation time, and the actual speedup gain can be observed: around 212 times in this case.

\begin{figure}[htbp]
    \centering
    \subfloat{%
    \resizebox*{6.5cm}{!}{\includegraphics{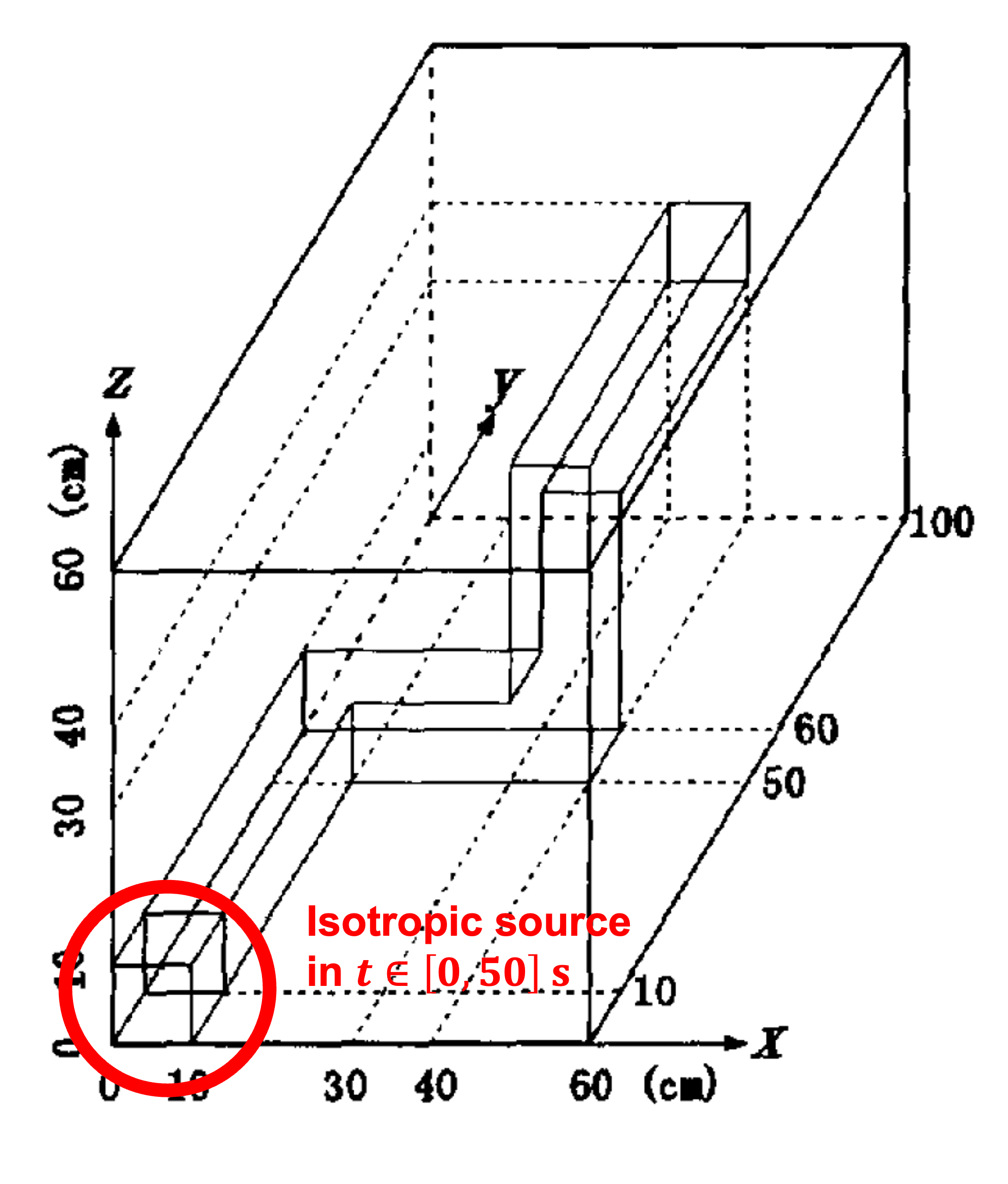}}}\hspace{8mm}
    \subfloat{%
    \resizebox*{5cm}{!}{\includegraphics{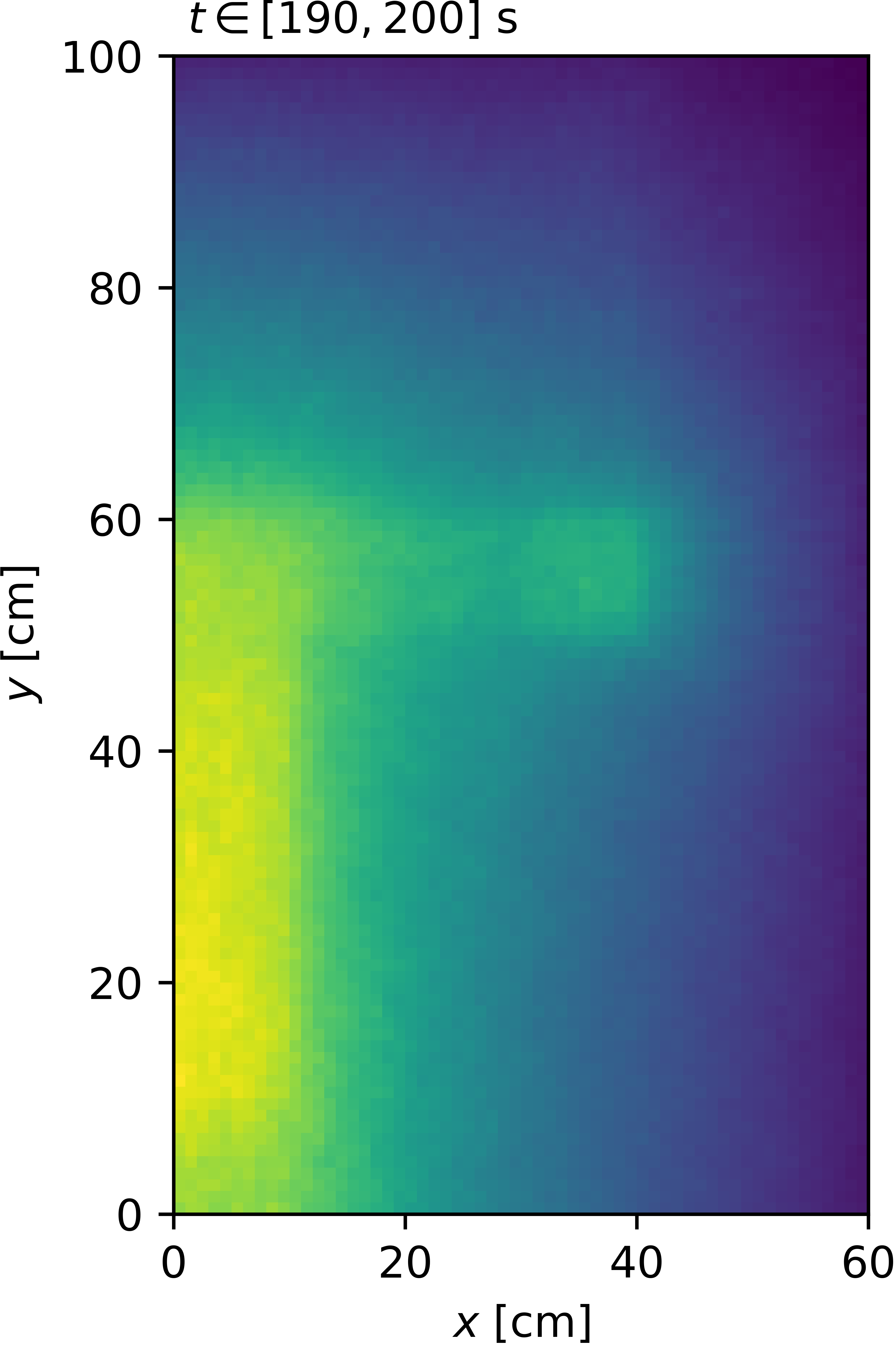}}}
    \caption{The time-dependent Kobayashi dog-leg void duct problem~\cite{kobayashi2001} (left) and the last snapshot of the normalized time-averaged flux distribution (10$^{\boldsymbol{8}}$ histories) (right).} \label{fig:kobayashi}
\end{figure}

To further exercise the time-dependent capabilities of MC/DC, we devised other transient problems based on existing benchmarks.
These include the three-dimensional dog-leg void duct Kobayashi radiation transport problem~\cite{kobayashi2001}, whose model is shown in the left side of Fig.~\ref{fig:kobayashi}.
We make the originally steady-state problem time-dependent by setting the monoenergetic neutron speed to one cm/s and making the specified source active only in the first fifty seconds of the simulation.
The quantity of interest is the time-averaged evolution (track-length estimator) of the flux distribution up to 200 seconds into the simulation.

Figure~\ref{fig:kobayashi} shows the last snapshot of the normalized flux distribution of the time-dependent Kobayashi problem, obtained with $10^8$ particle histories with implicit capture.
For this problem, Numba mode runs about 56 times faster than the Python mode, which is significantly lower than the number for the supercritical AZURV1 problem (212$\times$).
However, when we coarsen the spatial tally mesh of the Kobayashi problem by a factor of 10, the Numba mode runs about 309 times faster.
This sensitivity of Numba speedup to the size (and dimension) of the tally mesh may be due to how global variables (simulation model, parameters, and tallies) are currently managed in MC/DC: by using a single, large Numpy~\cite{harris2020array} structure passed around as a function argument at each transport kernel call.
We do this because Numba does not support global variables, as JIT compilation is done per each individual function.
Refactoring the large global variables container into smaller optimal chunks may further improve the Numba speedup.

\subsection{Code-to-Code Comparison and Parallel Scalability}

As an initial effort to compare the performance of the Python-based, JIT-compiled code MC/DC with conventional compiled MC codes, we set up a multigroup infinite pin cell criticality problem based on the C5G7-TD UO2 pin model~\cite{hou2017}.
We solve the $k$-eigenvalue problem by running 50 inactive and 100 active cycles with $10^5$ particles per cycle using MC/DC and the MC code Shift~\cite{PANDYA2016239} on LLNL’s compute platform Lassen (IBM POWER9).
MC/DC with Numba runs the problem in 2563.8 seconds, while Shift runs the problem in 1020.7 seconds.
Subtracting the Numba-JIT compilation time (which is around 63 seconds for this problem), Shift performs about 2.5 times faster than MC/DC, which is a respectable level for MC/DC considering the relatively low effort required to develop a Python-based code decorated by Numba's JIT.
However, if we also calculate the flux distribution in a fine mesh during the eigenvalue simulation, the performance of MC/DC considerably degrades.
With $20\times20\times100\times7$ and $100\times100\times500\times7$ mesh sizes (seven for the number of groups), Shift runs about 3 and 15 times faster than MC/DC, respectively.
This could be related to the suboptimal Numba speedup in problems with large-size multidimensional tallies discussed in the previous subsection.
However, this may also indicate that Shift has more optimized tallying mechanics/algorithms. 

\begin{figure}[htbp]
    \centering
    \subfloat{%
    \resizebox*{6.5cm}{!}{\includegraphics{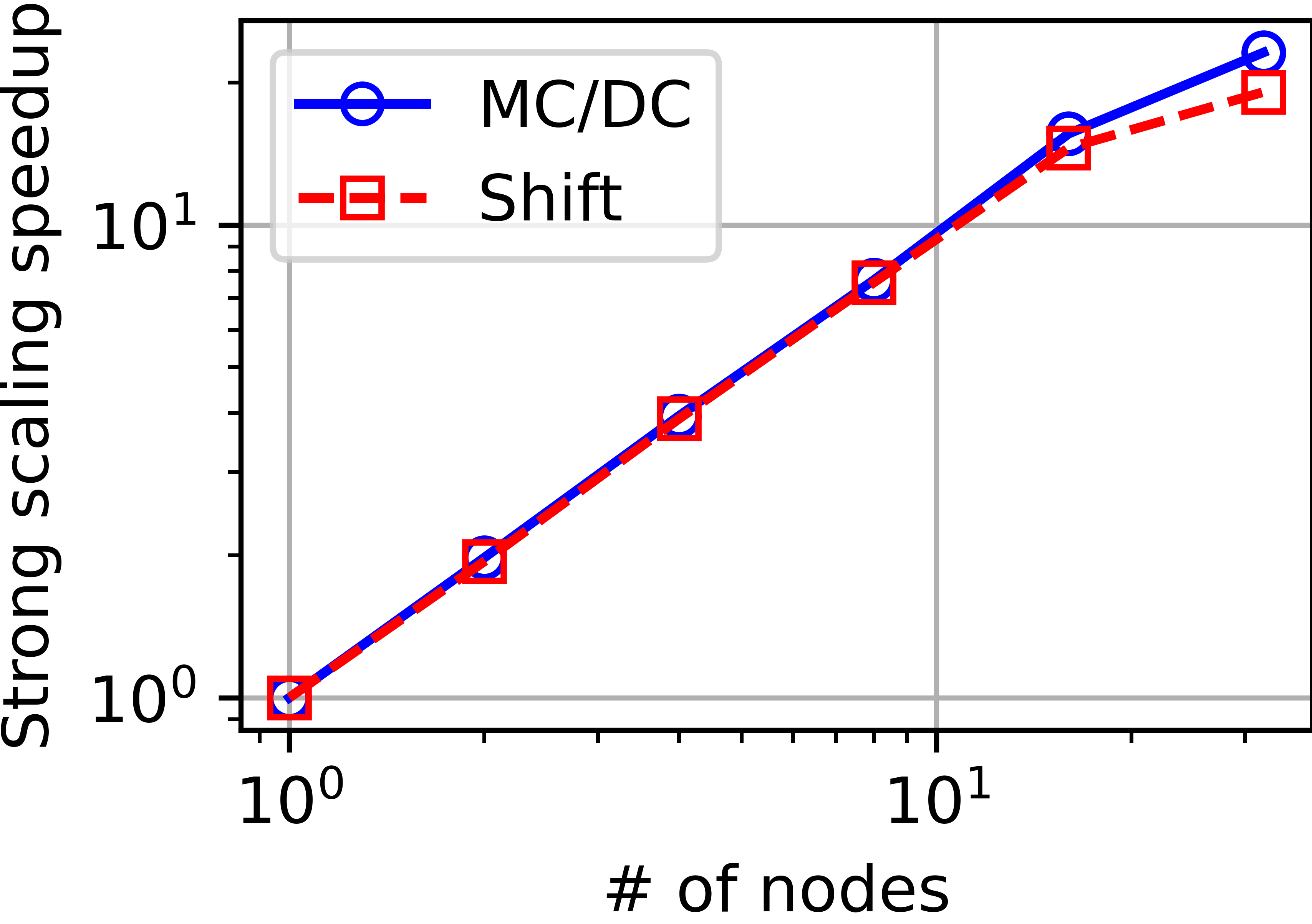}}}\hspace{9mm}
    \subfloat{%
    \resizebox*{6.5cm}{!}{\includegraphics{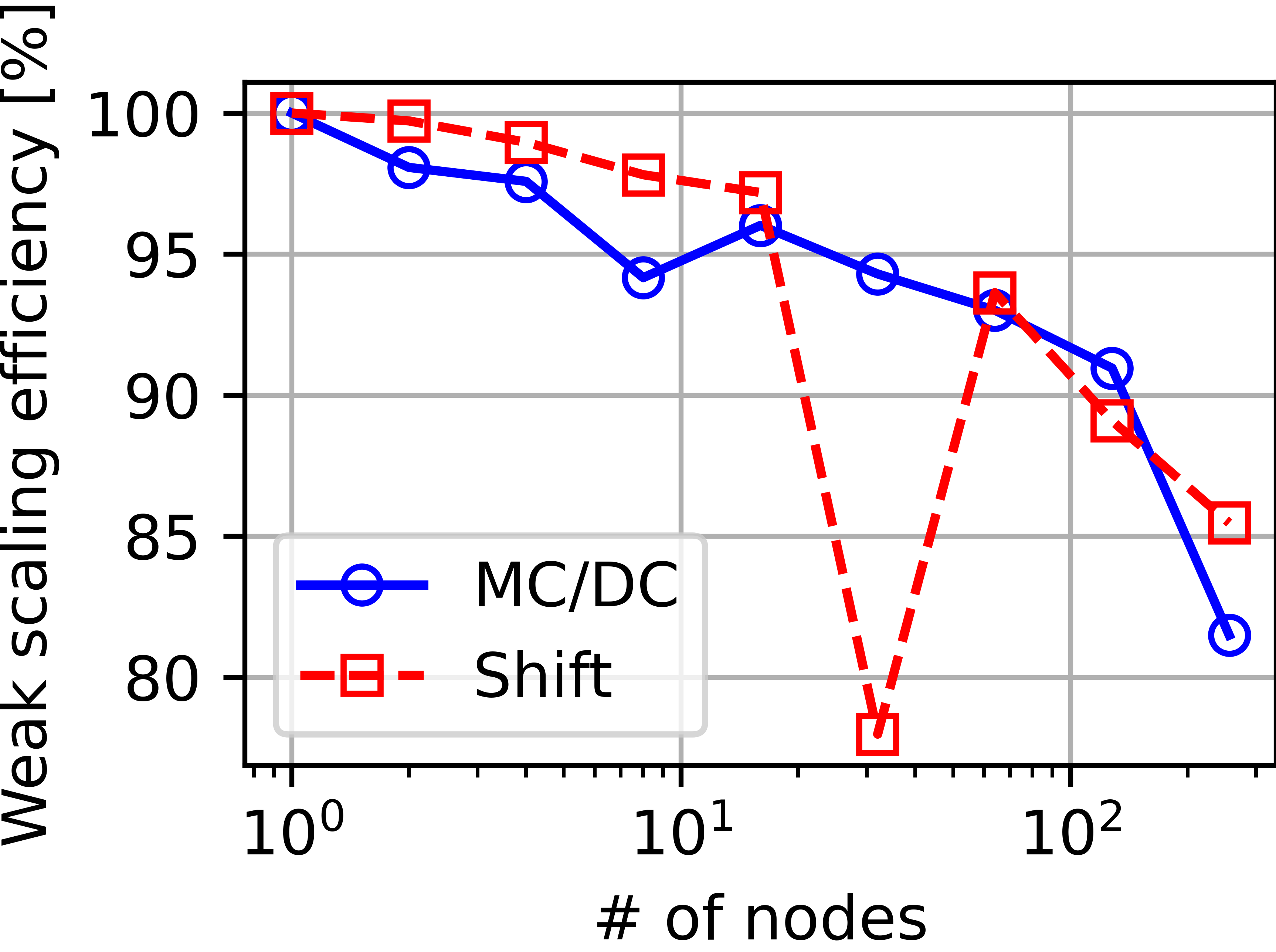}}}
    \caption{Strong (left) and weak scaling results (right) for the infinite UO2 pin cell multigroup criticality calculation. Shift and MC/DC were run on IBM POWER9 (40 cores/node) and Intel Xeon E5-2695 (36 cores/node) CPUs, respectively.} \label{fig:scaling}
\end{figure}

We also perform initial parallel scaling studies using the same infinite pin cell eigenvalue problem.
The Lassen architecture uses the specialized IBM Spectrum MPI, where MC/DC currently runs into MPI4Py issues when attempting to scale to large numbers of compute nodes.
Thus, for these initial scaling studies, MCDC uses LLNL's compute platform Quartz (Intel Xeon E5-2695). Figure~\ref{fig:scaling} shows the strong and weak scaling results. We run 25000 particles/cycle per core for the weak scaling; as for the strong scaling, we run with $25000\times 36$ and $25000\times 40$ for MC/DC and Shift, respectively. The results show that MC/DC scales comparably with Shift for the cases considered.

\subsection{C5G7-TD Benchmark and Challenge Problem}

\begin{figure}[htbp]
  \centering
  \includegraphics[scale=0.8]{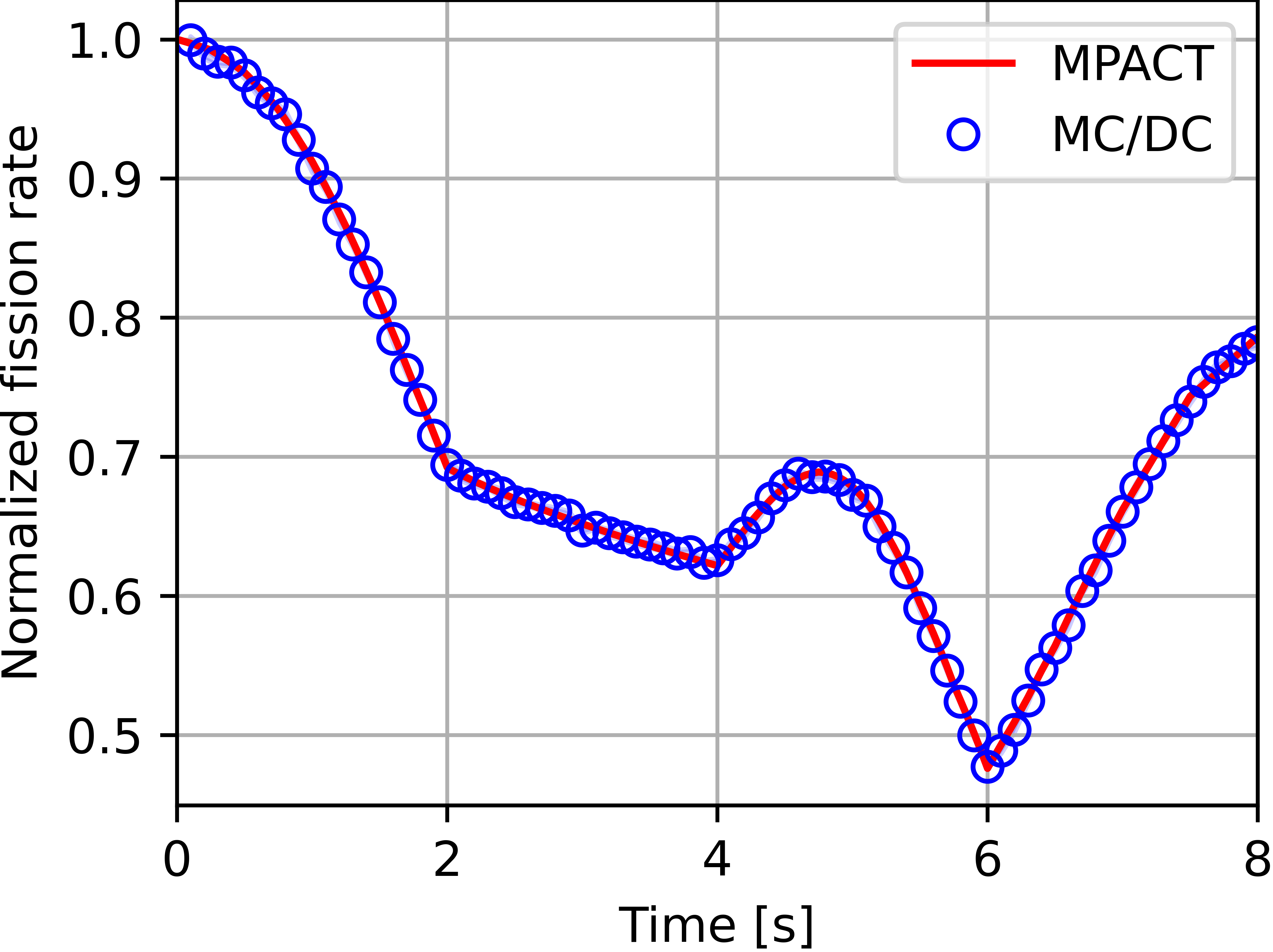}
  \caption{MC/DC result for the 3D C5G7 TD4-4 benchmark problem (eight batches of $\boldsymbol{5 \times 10^6}$ initial prompt and delayed neutrons, with time-crossing estimator). MPACT result~\cite{shen2019} is also presented as a reference solution.}
  \label{fig:c5g7-td}
\end{figure}

As the last verification study in this initial phase of MC/DC development, we consider one of the more involved multigroup 3D C5G7-TD benchmark problems: the exercise TD4-4~\cite{hou2017}.
The four-assembly problem involves continuous, overlapping insertion and withdrawal of two control rod banks.
To run a MC simulation of this type of reactor transient (i.e., starting from a critical steady state), we need to prepare the initial-condition neutrons and delayed neutron precursors.
We use the initial particle sampling technique proposed by Variansyah and McClarren~\cite{variansyah2023mc_ic}.

First, we run an accurate criticality calculation on the un-rodded configuration: with 50 inactive and 150 active cycles and 20 million particles per cycle, we get a $k$-eigenvalue of $1.165366 \pm 2.8$ pcm.
We then prepare the initial-condition particles by setting the neutron and delayed neutron precursor target sizes to be $5 \times 10^6$ \cite{variansyah2023mc_ic}. 
The problem is run in ``analog'' (uniform weight, without any variance reduction technique or population control).
The total fission rate is recorded via the time-crossing estimator in a uniform time grid of $\Delta t=0.1$ s.
Finally, the time-dependent surface~\cite{variansyah2023mc_tdSurf} is applied to exactly model the continuous movement of the control rod banks.
The analog transient MC simulation takes about 6 hours on 9216 cores of LLNL's compute platform Quartz.
The time-dependent simulation (and its initial condition generation) is repeated eight times with different random number seeds to measure the result's uncertainty. 
Figure~\ref{fig:c5g7-td} shows the result, which agrees well with the result generated by Shen et al. using the deterministic code MPACT~\cite{shen2019}.

\begin{figure}[htbp]
    \centering
    \subfloat{%
    \resizebox*{7cm}{!}{\includegraphics{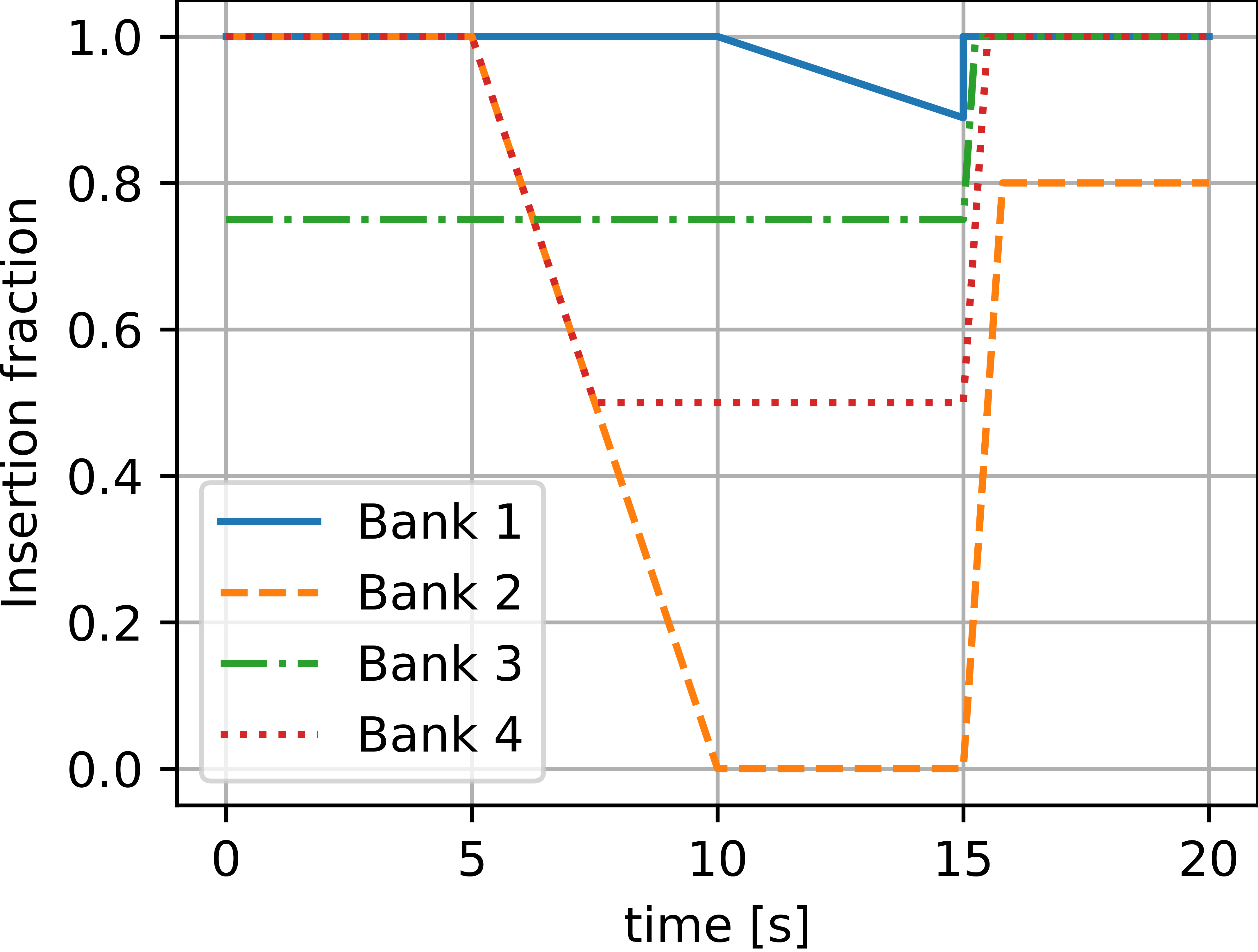}}}\hspace{5pt}
    \subfloat{%
    \resizebox*{7cm}{!}{\includegraphics{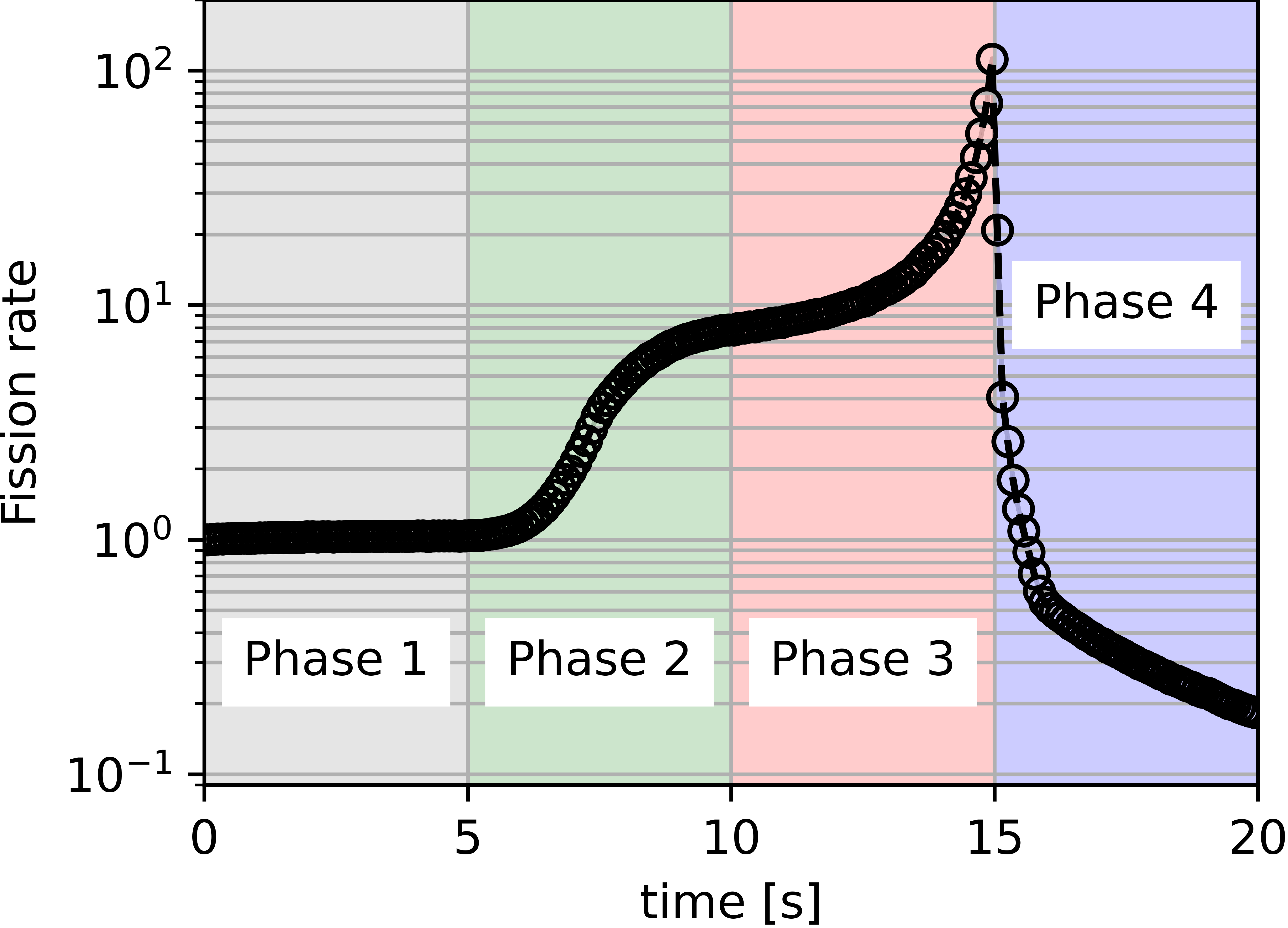}}}
    \caption{Control rod positions (left) and MC/DC solution (right) of C5G7-TDX.} \label{fig:c5g7tdx}
\end{figure}

Last but not least, part of MC/DC development under CEMeNT's project is developing challenge problems to measure the effectiveness of algorithms and methods being developed.
One of these is based on the 3D C5G7-TD~\cite{hou2017} problem, referred to here as C5G7-TDX.
Here, we devise a four-phase problem simulating a start-up accident experiment, where each phase lasts five seconds.
The problem is driven by a fixed source (in the fastest group) isotropically emitting neutrons at the center of Bank 1.
In Phase 1, all control rod banks are fully inserted, except Bank 3, which is stuck at 0.74 insertion fraction; this phase simulates source particle propagation to a steady state in a sub-critical system.
In Phase 2, control rod Banks 2 and 4 are fully withdrawn within five seconds, but Bank 4 is stuck halfway; this phase exhibits the typical S-wave control rod worth shape as the fission rate rises.
In Phase 3, Bank 1 slowly withdraws up to the insertion fraction of 0.889, which is enough to induce a neutron excursion.
In Phase 4, the fixed source is removed, and reactor scram is initiated, where all control rod banks drop at 0.1 insertion fractions per second, except for Bank 2, which gets stuck at the insertion fraction of 0.8; in this phase, rapid neutron population collapse occurs, but the decay rate is limited by the contributions of the delayed neutron precursors accumulated during the previous phases.
Figure~\ref{fig:c5g7tdx} (left) shows the control rod banks' positions during the four-phase simulation, while MC/DC total time-averaged fission rate solution from running $10^9$ analog histories is shown on the right.

\begin{figure}[htbp]
  \centering
  \includegraphics[scale=0.7]{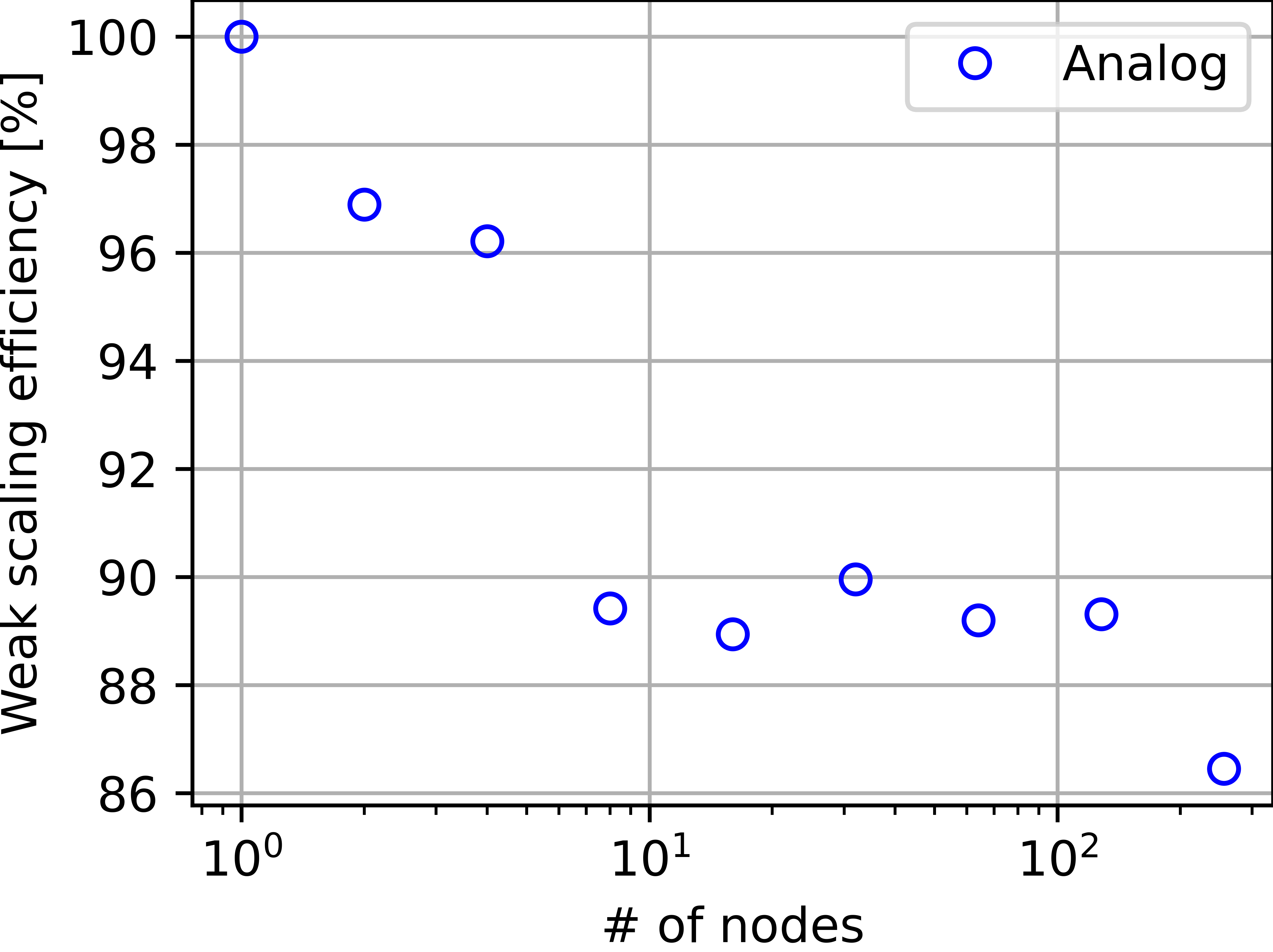}
  \caption{MC/DC parallel weak scaling efficiency (10$^{\boldsymbol{6}}$ histories/core and 36 cores/node) for C5G7-TDX.}
  \label{fig:c5g7tdx_scale}
\end{figure}

Figure~\ref{fig:c5g7tdx_scale} shows weak parallel scaling of C5G7-TDX for up to 256 nodes (36 cores per node) with $10^6$ histories per core.
Despite the relatively high number of histories per core, the efficiency still significantly decreases as we increase the number of nodes.
This indicates a significant load imbalance in the analog MC simulation of C5G7-TDX and makes it a reasonable challenge problem.

\section{SUMMARY AND FUTURE WORK}\label{sec:future}

MC/DC is designed to serve as a Python-based tool for MC transport method and algorithm explorations.
By leveraging MPI4Py and Numba JIT compilation libraries, as well as innovative abstraction strategy (Fig.~\ref{fig:adapter}) and compilation scheme (Fig.~\ref{fig:compiler}), MC/DC seeks to offer improved performance (beyond what is achievable by typical pure Python exploratory tools), massive scalability, and backend portability.

Here, we verified MC/DC's current capabilities against several test and benchmark problems.
We also proposed a challenge problem based on the C5G7-TD benchmark model to further exercise MC/DC time-dependent features.
Initial performance assessment shows that MC/DC's Numba mode can run hundreds of times faster than the pure Python mode. 
However, the speedup degrades if we significantly increase the size of the multidimensional tally, which is an opportunity to improve the performance further by optimizing the current MC/DC's Numba implementation.
Furthermore, running a simple infinite lattice eigenvalue problem showed that MC/DC runs about 2.5 to 3 times slower than the MC code Shift; however, the runtime ratio grows as we increase the mesh tally complexity, which further warrants optimization of MC/DC's Numba implementation.
Nevertheless, we demonstrated that MC/DC manages to match the excellent parallel scalability of Shift for this simple problem.

Currently, MC/DC only supports running multigroup transport problems on CPUs with the typical history-based MC algorithm.
Future work includes developing continuous energy physics capabilities and implementing the proposed abstraction strategy (Fig.~\ref{fig:adapter}) and compilation scheme (Fig.~\ref{fig:compiler}), which are currently being investigated in a light version of the code.

\section*{ACKNOWLEDGEMENTS}
This work was supported by the Center for Exascale Monte-Carlo Neutron Transport (CEMeNT), a PSAAP-III project funded by the Department of Energy, grant number DE-NA003967.

\setlength{\baselineskip}{12pt}
\bibliographystyle{mc2023}
\bibliography{mc2023}

\begin{thebibliography}{10}
\newcommand{\enquote}[1]{``#1''}
\providecommand{\url}[1]{\texttt{#1}}
\providecommand{\urlprefix}{URL }

\bibitem{zenodoMCDC}
I.~Variansyah, J.~P. Morgan, C.~Goodman, S.~Pasmann, and J.~Northrop.
\newblock \enquote{Monte Carlo / Dynamic Code (MC/DC).} (2023).
\newblock \urlprefix\url{https://doi.org/10.5281/zenodo.7916215}.
\newblock Version 0.1.0.

\bibitem{Witherden2014PyFR}
F.~D. Witherden, A.~M. Farrington, and P.~E. Vincent.
\newblock \enquote{{PyFR: An open source framework for solving
  advection-diffusion type problems on streaming architectures using the flux
  reconstruction approach}.}
\newblock \emph{Computer Physics Communications}, \textbf{volume 185}(11), pp.
  3028--3040 (2014).
\newblock \urlprefix\url{http://dx.doi.org/10.1016/j.cpc.2014.07.011}.

\bibitem{morgan2022}
J.~P. Morgan, T.~S. Palmer, and K.~E. Niemeyer.
\newblock \enquote{Explorations of Python-Based Automatic Hardware Code
  Generation for Neutron Transport Applications.}
\newblock In \emph{Transactions of the American Nuclear Society}, volume 126,
  pp. 318--320. Anaheim, CA, USA (2022).

\bibitem{lamNumba2015}
S.~K. Lam, A.~Pitrou, and S.~Seibert.
\newblock \enquote{Numba: A LLVM-Based Python JIT Compiler.}
\newblock In \emph{Proceedings of the Second Workshop on the LLVM Compiler
  Infrastructure in HPC}, LLVM '15. Association for Computing Machinery, New
  York, NY, USA.
\newblock \urlprefix\url{https://doi.org/10.1145/2833157.2833162}.

\bibitem{llvm}
\enquote{{The LLVM Compiler Infrastructure}.} (2022).
\newblock \urlprefix\url{https://llvm.org/}.

\bibitem{cuneothesis}
B.~Cuneo.
\newblock \emph{Divergence Reduction and Dependency Management in GPU Programs
  using Asynchronous Work Scheduling}.
\newblock Ph.D. thesis, Corvallis, OR (2022).

\bibitem{DALCIN20051108}
L.~Dalcín, R.~Paz, and M.~Storti.
\newblock \enquote{MPI for Python.}
\newblock \emph{Journal of Parallel and Distributed Computing},
  \textbf{volume~65}(9), pp. 1108--1115 (2005).
\newblock
  \urlprefix\url{https://www.sciencedirect.com/science/article/pii/S0743731505000560}.

\bibitem{zenodoMCDC-BIT}
I.~Variansyah, J.~P. Morgan, and B.~Cuneo.
\newblock \enquote{MC/DC - Back in Track.} (2023).
\newblock \urlprefix\url{https://doi.org/10.5281/zenodo.7916258}.

\bibitem{variansyah2022pct}
I.~Variansyah and R.~G. McClarren.
\newblock \enquote{Analysis of Population Control Techniques for Time-Dependent
  and Eigenvalue Monte Carlo Neutron Transport Calculations.}
\newblock \emph{Nuclear Science and Engineering}, \textbf{volume 196:11}, pp.
  1280--1305 (2022).

\bibitem{variansyah2022physor}
I.~Variansyah and R.~G. McClarren.
\newblock \enquote{{Performance of Population Control Techniques in Monte Carlo
  Reactor Criticality Simulations}.}
\newblock In \emph{Proc. PHYSOR 2022}. American Nuclear Society (2022).

\bibitem{variansyah2023mc_tdSurf}
I.~Variansyah and R.~G. McClarren.
\newblock \enquote{{High-fidelity treatment for object movement in
  time-dependent Monte Carlo transport simulations}.}
\newblock In \emph{Proc. M\&C 2023}. American Nuclear Society (2023).

\bibitem{variansyah2023mc_ic}
I.~Variansyah and R.~G. McClarren.
\newblock \enquote{{An effective initial particle sampling technique for Monte
  Carlo reactor transient simulations}.}
\newblock In \emph{Proc. M\&C 2023}. American Nuclear Society (2023).

\bibitem{Ganapol2001HomogeneousBenchmarks}
B.~Ganapol, R.~Baker, J.~Dahl, and R.~E. Alcouffe.
\newblock \enquote{Homogeneous Infinite Media Time-Dependent Analytical
  Benchmarks.}
\newblock In \emph{International Meeting on Mathematical Methods for Nuclear
  Applications}, volume 41(25). Salt Lake City, UT (2001).

\bibitem{kobayashi2001}
K.~Kobayashi and N.~Sugimara.
\newblock \enquote{3D Radiation Transport Benchmark Problems and Results for
  Simple Geometries with Void Region.}
\newblock \emph{Progress in Nuclear Energy}, \textbf{volume 39:2}, pp. 119--144
  (2001).

\bibitem{harris2020array}
C.~R. Harris, K.~J. Millman, S.~J. van~der Walt, R.~Gommers, P.~Virtanen,
  D.~Cournapeau, E.~Wieser, J.~Taylor, S.~Berg, N.~J. Smith, R.~Kern, M.~Picus,
  S.~Hoyer, M.~H. van Kerkwijk, M.~Brett, A.~Haldane, J.~F. del R{\'{i}}o,
  M.~Wiebe, P.~Peterson, P.~G{\'{e}}rard-Marchant, K.~Sheppard, T.~Reddy,
  W.~Weckesser, H.~Abbasi, C.~Gohlke, and T.~E. Oliphant.
\newblock \enquote{Array programming with {NumPy}.}
\newblock \emph{Nature}, \textbf{volume 585}(7825), pp. 357--362 (2020).
\newblock \urlprefix\url{https://doi.org/10.1038/s41586-020-2649-2}.

\bibitem{hou2017}
J.~J. Hou, K.~N. Ivanov, V.~F. Boyarinov, and P.~A. Fomichenko.
\newblock \enquote{{OECD/NEA benchmark for time-dependent neutron transport
  calculations without spatial homogenization}.}
\newblock \emph{Nuclear Engineering and Design}, \textbf{volume 317}, pp.
  177--189 (2017).

\bibitem{PANDYA2016239}
T.~M. Pandya, S.~R. Johnson, T.~M. Evans, G.~G. Davidson, S.~P. Hamilton, and
  A.~T. Godfrey.
\newblock \enquote{Implementation, capabilities, and benchmarking of Shift, a
  massively parallel Monte Carlo radiation transport code.}
\newblock \emph{Journal of Computational Physics}, \textbf{volume 308}, pp.
  239--272 (2016).

\bibitem{shen2019}
Q.~Shen, Y.~Wang, D.~Jabaay, B.~Kochunas, and T.~Downar.
\newblock \enquote{{Transient analysis of C5G7-TD benchmark with MPACT}.}
\newblock \emph{Annals of Nuclear Energy}, \textbf{volume 125}, pp. 107--120
  (2019).

\end{thebibliography}
\end{document}